\documentclass[9pt]{article}
\usepackage{xcolor}
\usepackage{balance}

\usepackage{cite}
\usepackage{amsmath,amssymb,amsfonts}
\usepackage{algorithmic}
\usepackage{graphicx}
\usepackage{textcomp}
\usepackage{array}
\usepackage{multirow}

\usepackage{steinmetz}

\begin{document}
%
\title{Super-Realized Gain Huygens Antennas}




\author{
\small
    Donal Patrick Lynch$^*$ \and
    \small
    Vincent Fusco\thanks{Electronics Communications and Information Technology, ECIT, Queen’s University Belfast, Belfast, U.K.; Email: \texttt{{dlynch27@qub.ac.uk, v.fusco@ecit.qub.ac.uk, s.asimonis@qub.ac.uk}}} \and
    \small
    Manos M. Tentzeris\thanks{Georgia Institute of Technology, School of Electrical and Computer Engineering, Atlanta, GA 30332, USA; Email: \texttt{etentze@ece.gatech.edu}} \and
    \small
    Stylianos D. Asimonis$^*$}

\date{} %
\maketitle

\begin{abstract}
 This study presents a superdirective antenna array specifically designed for the sub-6 5G frequency range, incorporating pioneering Huygens antenna elements. The optimized structure achieves a realized gain that surpasses Harrington's well-known maximum theoretical limit for antenna directivity, effectively addressing practical concerns related to ohmic and return losses. Additionally, the compact size of the proposed antenna array, as opposed to a uniform linear array, offers the potential to fabricate highly radiation-efficient, high-directivity antennas in a compact form.
\end{abstract}


\section{Introduction}

Antenna arrays play a crucial role in enhancing antenna performance, particularly in achieving higher directivity. The traditional approach to enhance the directivity of a uniformly excited array often requires utilizing a large number of elements, leading to undesirably large and high-power consumption communication systems. In \cite{uzkov}, Uzkov theoretically demonstrated that a uniform linear array (ULA) consisting of $N$ isotropic elements with extremely small inter-element spacing and non-uniform excitation signals exhibits an end-fire directivity of $N^2$. This phenomenon is widely recognized as \textit{superdirectivity}. Therefore, a superdirective array surpasses the maximum directivity of a ULA (i.e., $N$) by meticulously adjusting the magnitude and phase of the voltage signals driving the radiating elements \cite{hansen,altshuler,Dovelos2023,best} and by placing the elements very close together. As a result, superdirective antenna arrays have gained significant interest as a potential solution for designing directional antenna systems. However, this close proximity between the radiating elements results in strong coupling, leading to high corresponding input currents. These currents induce significant ohmic losses that adversely affect both radiation efficiency and antenna gain \cite{Dovelos2022, Lynch2023}. Additionally, superdirectional antenna arrays tend to exhibit high reactance in their input impedance characteristics due to this strong coupling. This complexity complicates efforts to achieve conventional $50, \Omega$ impedance matching, further diminishing the realized gain obtained.

In addition, a Huygens antenna is an array of two closely spaced elements, usually an electric dipole and a magnetic dipole, where the phases and amplitudes of the elements are precisely controlled to achieve highly directional radiation patterns \cite{debard18,ziolkowski}.

\begin{figure}[!t] 
\centering
\includegraphics[height=0.35\linewidth]{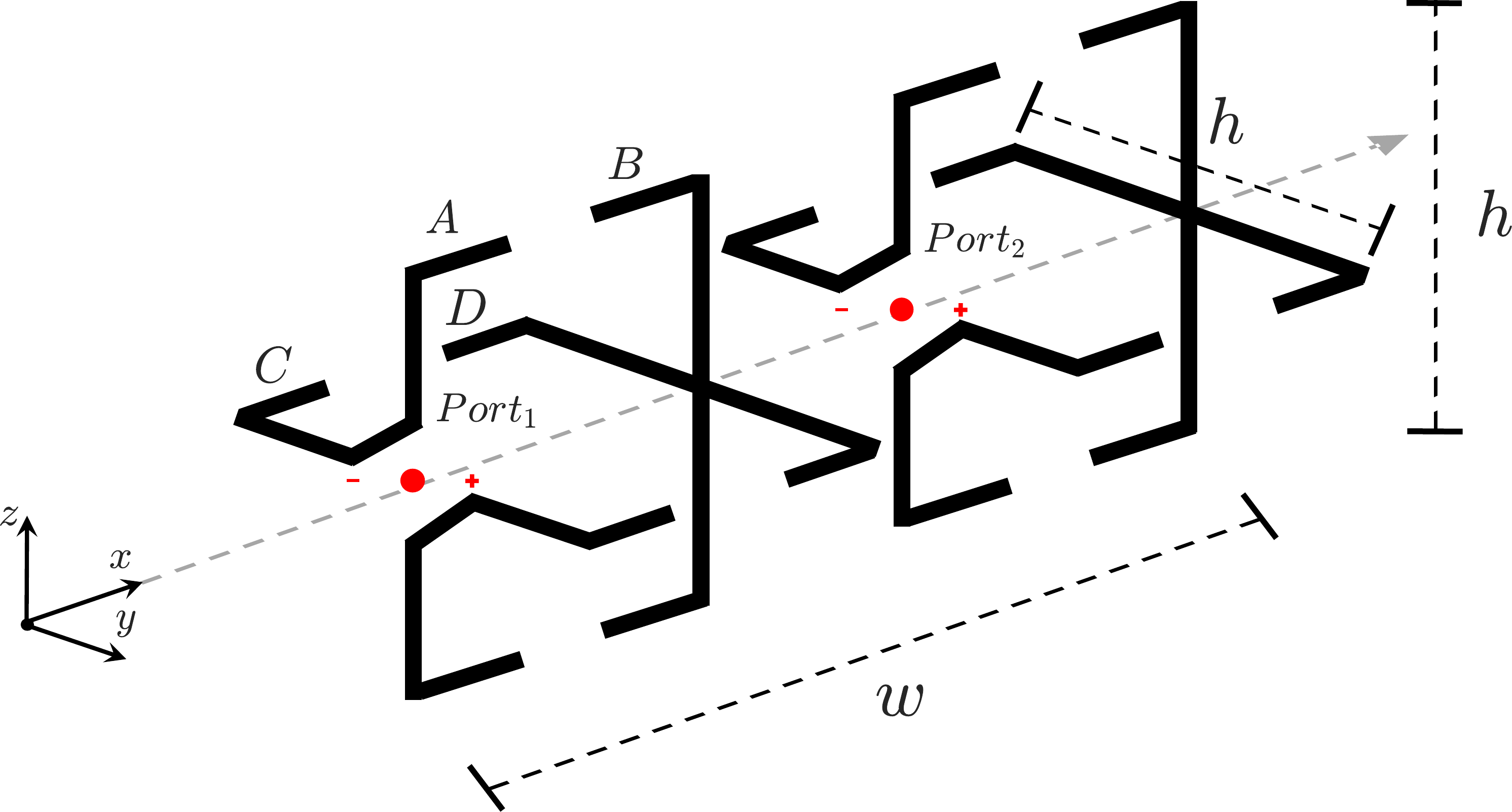}
\caption{A linear array of two QLA elements arranged in the $+x$ direction. Both are driven by signals $v_1$ and $v_2$ applied to port 1 and 2, respectively.}
\label{Fig1}
\end{figure}

\begin{figure}[!t] 
\centering
\includegraphics[height=0.28\linewidth]{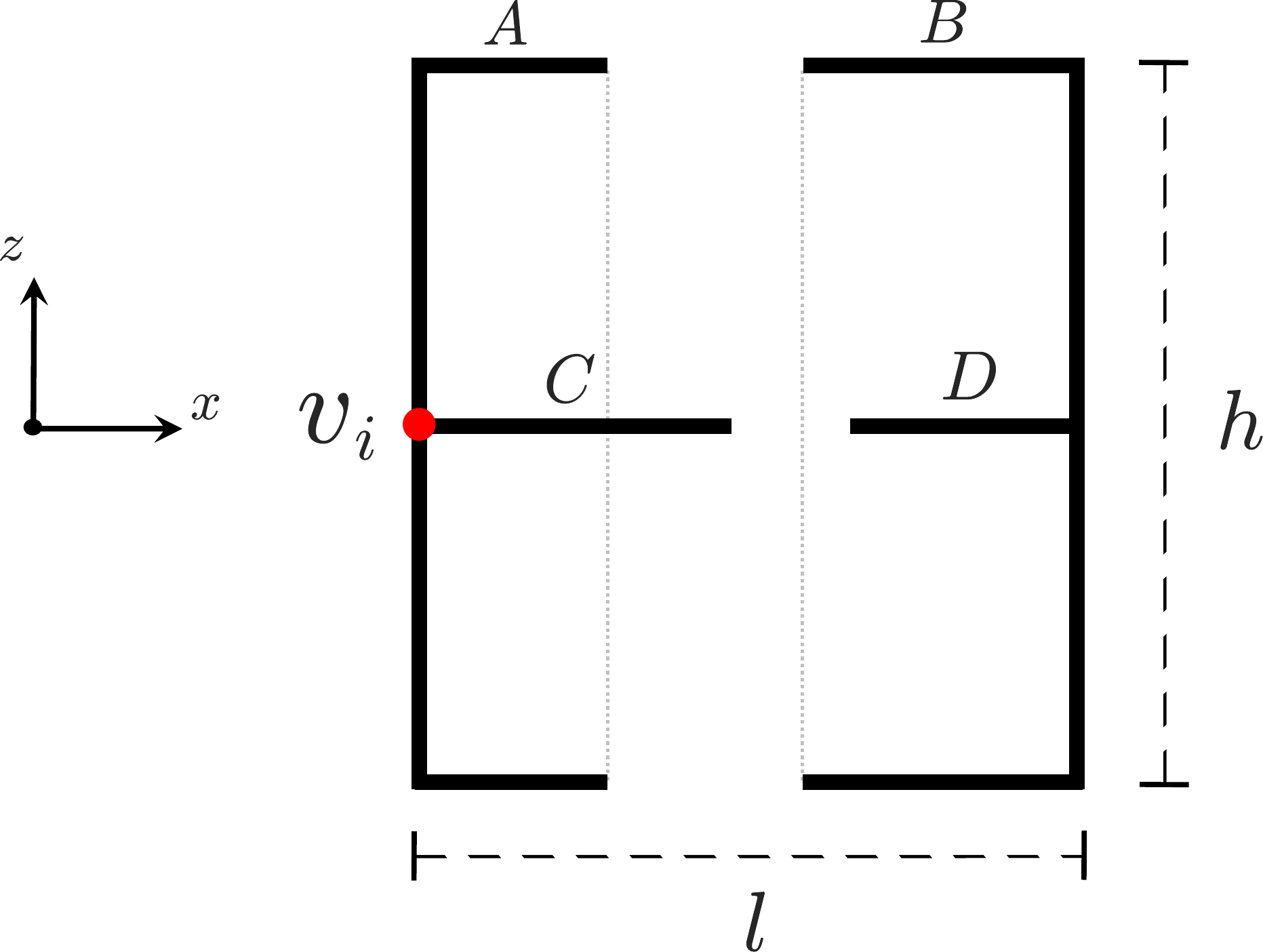}
\caption{Side view of a QLA element, with dimensions labeled as $A$, $B$, $C$, and $D$, and the driven voltage magnitude $v_i$ applied to the $i$-th ($i=1,2$) excitation port.}
\label{Fig2}
\end{figure}

The contribution of this work lies in the design of a super realized gain antenna system, achieving a realized gain higher than the theoretical directivity limit of $D=(k R)^2+2 k R$ proposed by Harrington in \cite{Harrington1958}, where $k$ represents the wavenumber and $R$ denotes the radius of a hypothetical sphere encapsulating the antenna system. In sharp contrast to the state-of-the-art, our focus is on the realized gain, which accounts for ohmic and return losses. To achieve this objective, we have incorporated Huygens radiating elements into the design process and maximized the realised gain of the antenna system as a function of the element dimensions and excitation signals. Our research aims to demonstrate the feasibility and effectiveness of this approach and its potential impact on various high-gain applications.

\section{Antenna Design}

This section presents the design of a two-element linear antenna array as shown in Fig. \ref{Fig1}. The antenna elements are based on the work of Fusco \textit{et al.} \cite{fusco}, which introduces the Quadrifilar Loop Antenna (QLA). A QLA antenna consists of two orthogonal split rectangular loops with different equivalent inductances and capacitances controlled by the length of the splits. 
%
%
It is important to note that the QLA was originally designed to produce circular polarization using these two orthogonally placed wire loops. Circularly polarized waves can be thought of as a combination of two linearly polarized waves with a $90^{\circ}$ degree phase shift. Consequently, the QLA generates wavefronts with both amplitude and phase variations, in accordance with the basic principles of the Huygens-Fresnel principle. In addition, the radiation pattern of the QLA has a cardioid shape, a key characteristic of Huygens antennas. Considering these factors, we consider the highly directive QLA to be a practical implementation of a Huygens antenna.
Both elements of the antenna array are excited individually with signals $v_1$ and $v_2$ respectively.


The commercial solver CST Studio Suite 2022 \cite{CST2022} was used to perform electromagnetic numerical simulations, specifically employing the integral equation solver. Our design focused on the fifth generation (5G) sub-6 (below 6 GHz) frequency range, particularly the 3.5 GHz band (i.e., $3.3-4.2$ GHz \cite{Eid2021}). The wires of the QLA elements were copper with a conductivity of $5.8 \times 10^7$ S/m and a diameter of $3$ mm. The size of each QLA element (Fig. \ref{Fig2}) was set to $h \times h \times l = 0.29\lambda \times 0.29\lambda \times 0.12\lambda$, where $\lambda$ represents the wavelength at $3.5$ GHz in free space. Additionally, the inter-element distance (i.e., from port 1 to port 2) was set to $0.17\lambda$. Consequently, the total size of the antenna array is $h \times h \times w = 0.29\lambda \times 0.29\lambda \times 0.28\lambda$ (or $25$~mm $\times$ $25$~mm $\times$ $24.28$~mm). Thus, the radius of the hypothetical sphere which encapsulates the proposed antenna system is $R=0.25\lambda$ (or $21.45$ mm), and the corresponding theoretical maximum achieved $D = (k R)^2 + 2 k R = 5.56$ or $7.5$ dBi. 

The antenna array was optimized to achieve an end-fire realized gain higher than the theoretical limit of $7.5$ dBi for directivity, subject to the dimensions $A$, $B$, $C$, and $D$, as well as the excitation signals $v_1$ and $v_2$ (Fig. \ref{Fig2}), in the $x$-direction. Mathematically speaking:

\begin{equation}
\begin{aligned}
\underset{
\left\lbrace A,B,C,D,v_1,v_2 \right\rbrace 
}{\text{Maximize}}
& \quad f \left(A,B,C,D,v_1,v_2\right) \\
\text{subj. to:} 
        & \quad A+B \le l,\\
        & \quad C+D \le l,\\
        & \quad |v_1|, |v_2| \in  \left[0, 1\right],\\
        & \quad \phase{v_1}=0,\\
        & \quad \phase{v_2} \in \left[0^{\circ}, 360^{\circ}\right],
\end{aligned}
\end{equation}
where $f$ denotes the realized gain function.

Particle Swarm Optimization (PSO) was employed as the optimization method. The optimized design parameters are presented in Table \ref{table01}, resulting in a maximum realized gain of $7.7$ dBi at $3.5$ GHz, surpassing Harrington's theoretical limit of $7.5$ dBi. The corresponding directivity is $8$ dBi. However, it is important to note that our optimization focused on realized gain rather than directivity because the latter does not consider antenna losses, making it a less practical property for real-world applications. Notably, realized gain accounts for both ohmic and return losses on the antenna, making it a more relevant antenna property for practical implementation.

\begin{table}[h]
\renewcommand{\arraystretch}{1.3}
\caption{Optimized Design Parameters at $3.5$ GHz}
\label{table_example}
\centering
\begin{tabular}{c|| 
>{\centering\arraybackslash}m{0.9cm} |
>{\centering\arraybackslash}m{0.9cm} |
>{\centering\arraybackslash}m{0.9cm} |
>{\centering\arraybackslash}m{0.9cm} |
>{\centering\arraybackslash}m{0.9cm} |
>{\centering\arraybackslash}m{0.9cm} }
\hline
Elem & $A$ (mm) & $B$ (mm) & $C$ (mm) & $D$ (mm) & $|v_i|$ (V) & $\phase{v_i}$ (deg.) \\
\hline\hline
1 & $5.1$ & $3.7$ & $5.3$ & $3.6$ & $0.14$ & 0\\
\hline
2 & $0.2$ & $6.2$ & $3.9$ & $4.7$ & $0.75$ & $153^\circ$\\
\hline
\end{tabular}
\label{table01}
\end{table}

\begin{figure}[t!] 
\centering
\includegraphics[height=0.4\linewidth]{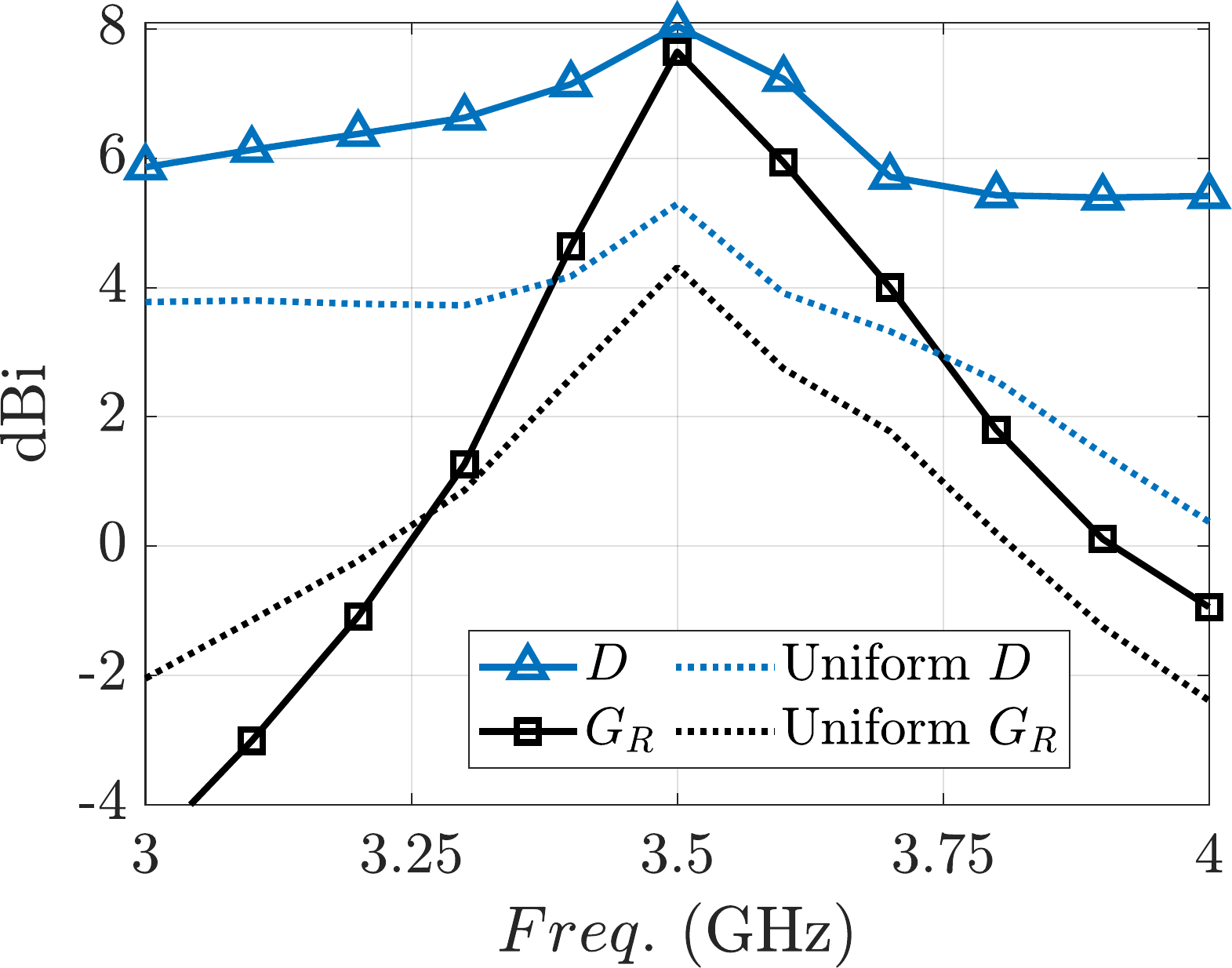}
\caption{The antenna array's directivity $D$ and realized gain $G_R$ for the optimized and uniform cases versus frequency from $3-4$ GHz.}
\label{Fig3}
\end{figure}

In Fig. \ref{Fig3}, the directivity and realized gain for both the optimal and uniform cases are plotted over the frequency range of $3 - 4$ GHz. In the uniform case, the elements are driven by voltages $v_1$ = $v_2$ = $1$ V, while all other dimensions are maintained identical to those in the optimized case (Table \ref{table01}). It is evident that the antenna array exhibits superdirectivity and super-realized gain in this frequency range \cite{hansen}.


\begin{figure}[!t] 
\centering
\includegraphics[height=0.35\linewidth]{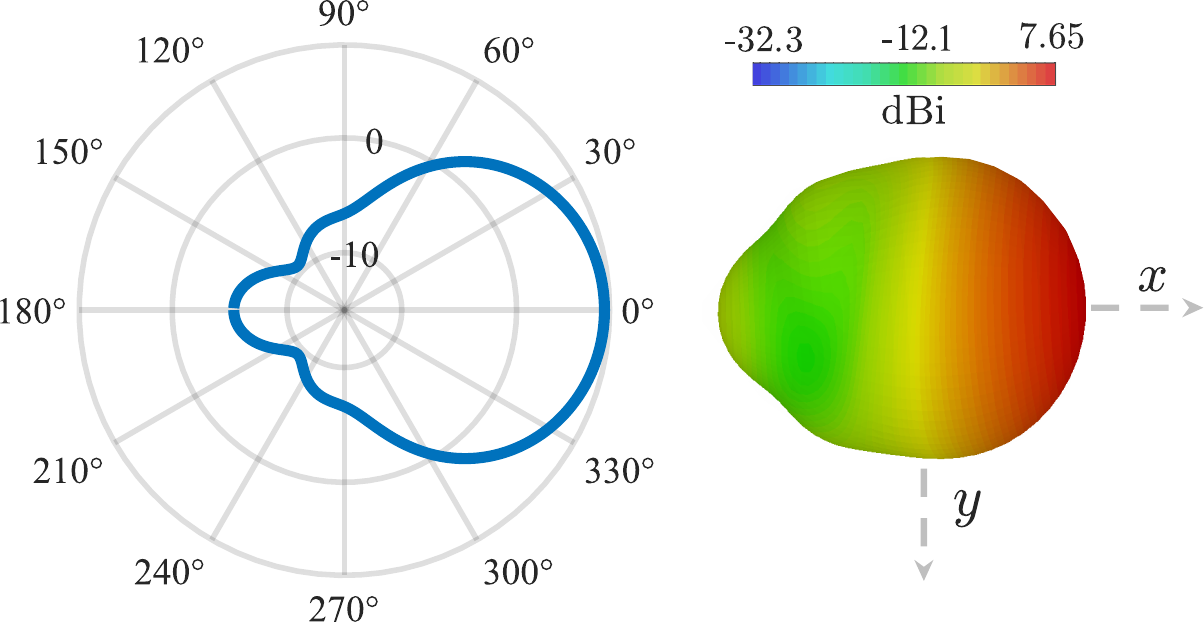}
\caption{The realized gain radiation pattern, horizontal plane ($E$-plane). Maximum occurs in $+x$ direction (i.e., end-fire antenna array).}
\label{Fig4}
\end{figure}

Fig. \ref{Fig4} depicts the simulated realized gain in the horizontal plane ($xy$-plane, Fig. \ref{Fig1}) for the optimal configuration. The achieved maximum realized gain is $7.7$ dBi, observed in the +$x$ direction, as expected, confirming an end-fire antenna array setup. The angular width, measured at the $3$ dB drop-off points, is $72^{\circ}$. Additionally, the front-to-back ratio at $3.5$ GHz measures $13.2$~dB.

\begin{figure}[!t] 
\centering
\includegraphics[height=0.4\linewidth]{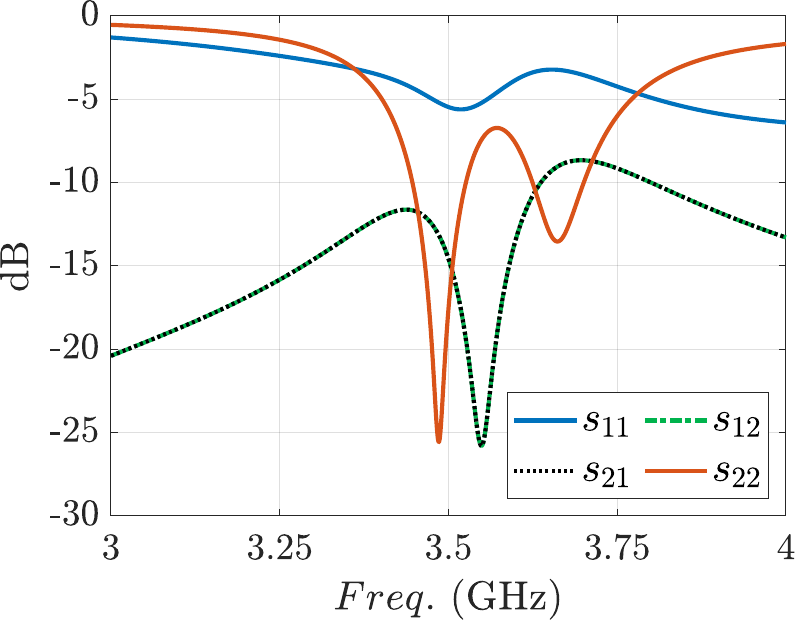}
\caption{Simulated S-parameters of the antenna array with a reference impedance of $50$ $\Omega$.}
\label{Fig5}
\end{figure}

\begin{figure}[!t] 
\centering
\includegraphics[height=0.4\linewidth]{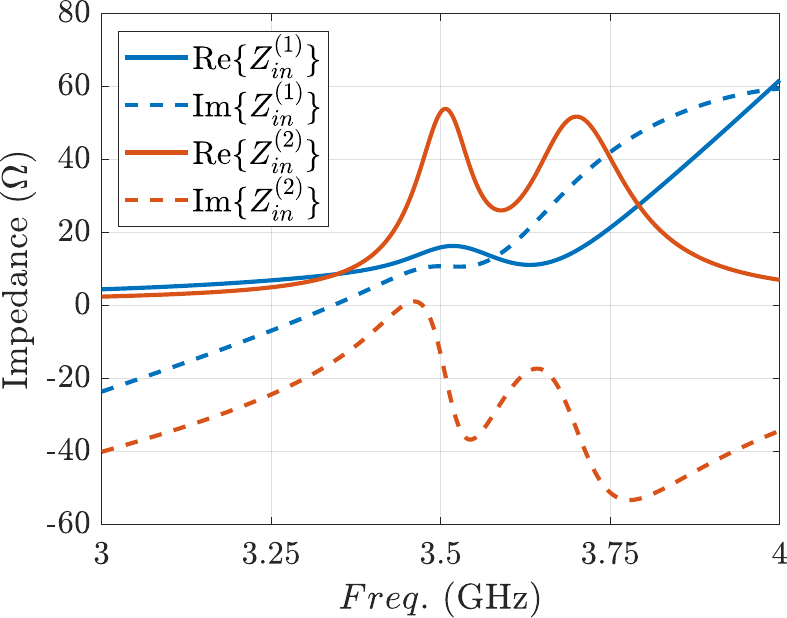}
\caption{Calculated impedance parameters of the antenna array elements.}
\label{Fig6}
\end{figure}

Fig. \ref{Fig5} illustrates the simulated $S$-parameters of the QLA array versus frequency at $50$ $\Omega$. At $3.5$ GHz, the reflection coefficient at port 1 and 2 is $-5.5$ dB and $-17.6$ dB, respectively. It is evident that element 2 is well impedance-matched, but the first element is not. This observation might raise concerns about the level of return losses. However, considering the voltage that drives elements 1 and 2 (Table \ref{table01}), it is noted that the power delivered to element 1 is low ($v_1$ is only 0.14 V) compared to the power delivered to element 2. Consequently, element 1 is acting as a parasitic element within the antenna array. Thus, the antenna system's realized gain remains high, surpassing Harrington's theoretical limit.
Fig. \ref{Fig6} illustrates the antenna impedance for both elements, and the results align with those shown in Fig. \ref{Fig5}. Specifically, the input impedance at port 1 and 2 is $16.9 + j10.9~\Omega$ and $53.1 - j13.4~\Omega $, respectively.

\section{Conclusion}
This research has explored the design and analysis of a simulated Huygens source antenna array, featuring two quadrifilar loop antenna elements which were optimized for realized gain, with inter-element distance of $0.17\lambda$. The study demonstrated that with careful design considerations, it is possible to achieve 
a superdirective array with remarkable isolation between closely spaced elements. This accomplishment opens up promising avenues for the practical deployment of compact, high-efficiency antenna arrays in densely populated wireless communication systems. As the demand for efficient and space-saving antenna solutions continues to grow in modern wireless communication, the insights gained from this research hold significant relevance and potential for future applications and advancements in the field.

\balance

\section*{Acknowledgment}

This work was funded by the Department of the Economy (DfE) in Northern Ireland. As part of their Postgraduate Studentships programme with Queen’s University Belfast.

\bibliographystyle{IEEEtran}
\bibliography{mybib.bib}

\end{document}